\documentclass[twocolumn,secnumarabic,amssymb, amsmath, tightenlines, nobibnotes, aps, prl, superscriptaddress]{revtex4-1}

\usepackage{bm}		
\usepackage{braket}
\usepackage{natbib}
\usepackage{amssymb,amsmath}
\usepackage{graphicx}

\usepackage[pdftex,colorlinks=true,pdfstartview=Fit]{hyperref}		

\usepackage{color}			

\newcommand{\av}[1]{\left\langle #1 \right\rangle}

\begin{document}

\title{Test of the diffusing-diffusivity mechanism using near-wall colloidal dynamics}
\author{Mpumelelo Matse}
\affiliation{Department of Physics, Simon Fraser University, Burnaby, B.C., V5A 1S6, Canada}

\author{Mykyta V. Chubynsky}
\altaffiliation[Present address: ]{Mathematics Institute, University of Warwick, Coventry CV4 7AL, United Kingdom}
\affiliation{Department of Physics, University of Ottawa, 150 Louis-Pasteur, Ottawa, Ontario K1N 6N5, Canada}

\author{John Bechhoefer}
\email[email: ]{johnb@sfu.ca}
\affiliation{Department of Physics, Simon Fraser University, Burnaby, B.C., V5A 1S6, Canada}

\begin{abstract}
The mechanism of diffusing diffusivity predicts that, in environments where the diffusivity changes gradually, the displacement distribution becomes non-Gaussian, even though the mean-squared displacement (MSD) grows linearly with  time.  Here, we report single-particle tracking measurements of the diffusion of colloidal  spheres near a planar substrate.  Because the local effective diffusivity is known, we have been able to carry out the first direct test of this mechanism for diffusion in inhomogeneous media.
\end{abstract}

\maketitle

The simple picture of Brownian motion due to Einstein, von Smoluchowski, and others \cite{Einstein,Smoluchowski,gard} leads to stochastic motion where the mean-square displacement (MSD) is linear in time $t$ and where displacements are Gaussian distributed.  This picture, appropriate for an isolated object diffusing in a homogeneous and infinite medium, breaks down in more complex environments.  For example, in some situations, the motion of molecules that diffuse inside the crowded environment of a cell has been described by anomalous diffusion, with an MSD having sublinear behavior $\sim t^\alpha$, with $0<\alpha<1$ \cite{an1,an2,an3,an4, an5}.  In some cases (described by the fractional-Brownian-motion model), the accompanying displacement distributions remain Gaussian \cite{anG1}, while in others (continuous-time-random-walk model)  they are non-Gaussian \cite{anNG1, anNG2}.

Another group of experiments has also reported  deviations from simple Brownian motion in complex media, with MSDs that are linear but with displacement distributions having tails that decay more slowly than Gaussian.  Such \textit{non-Gaussian yet normal} diffusion has been reported on lipid tubules and in networks of filamentous molecules \cite{NG1, NG2}, polymer systems \cite{NG3,banks16,xue16}, porous media \cite{NG5}, active-matter systems \cite{NG6,NG7}, supercooled liquids \cite{NG8,NG9}, and colloidal suspensions \cite{NG10}, as well as in simulations of 2D disks \cite{kim13} and porous media \cite{NG11}.  This behavior is believed to arise in complex environments where the effective diffusion constant varies in space.  The picture is that the observed motion is the superposition of ordinary diffusion processes that, over short time intervals, are simple, with Gaussian displacements characterized by some local diffusion constant.  Displacement distributions for an ensemble of particles then convolute the contribution of Gaussian distributions with different variances, sometimes leading to an overall non-Gaussian distribution.  At the same time, the central limit theorem ensures that longer-time displacements are Gaussian, with linearly increasing MSD.  This picture, developed qualitatively in \cite{NG1, NG2} and explained theoretically in \cite{NG12}, has inspired much theoretical investigation \cite{wang16,jain16a,jain16b,cherstvy16,samanta16,chechkin17}.

In the experiments done to date, the characteristics of the complex environment, such as the local value of $D$, were not known, except perhaps statistically.  Here, we report experimental observations of single-particle diffusion in a system where the underlying $D$ variations are known independently.  Knowing explicitly the variations in $D$, we then carry out the first direct test of the diffusing-diffusivity mechanism proposed in \cite{NG12} to account for non-Gaussian yet normal diffusion.  

\begin{figure}
	 \begin{center}
	 \includegraphics[width=8.4cm]{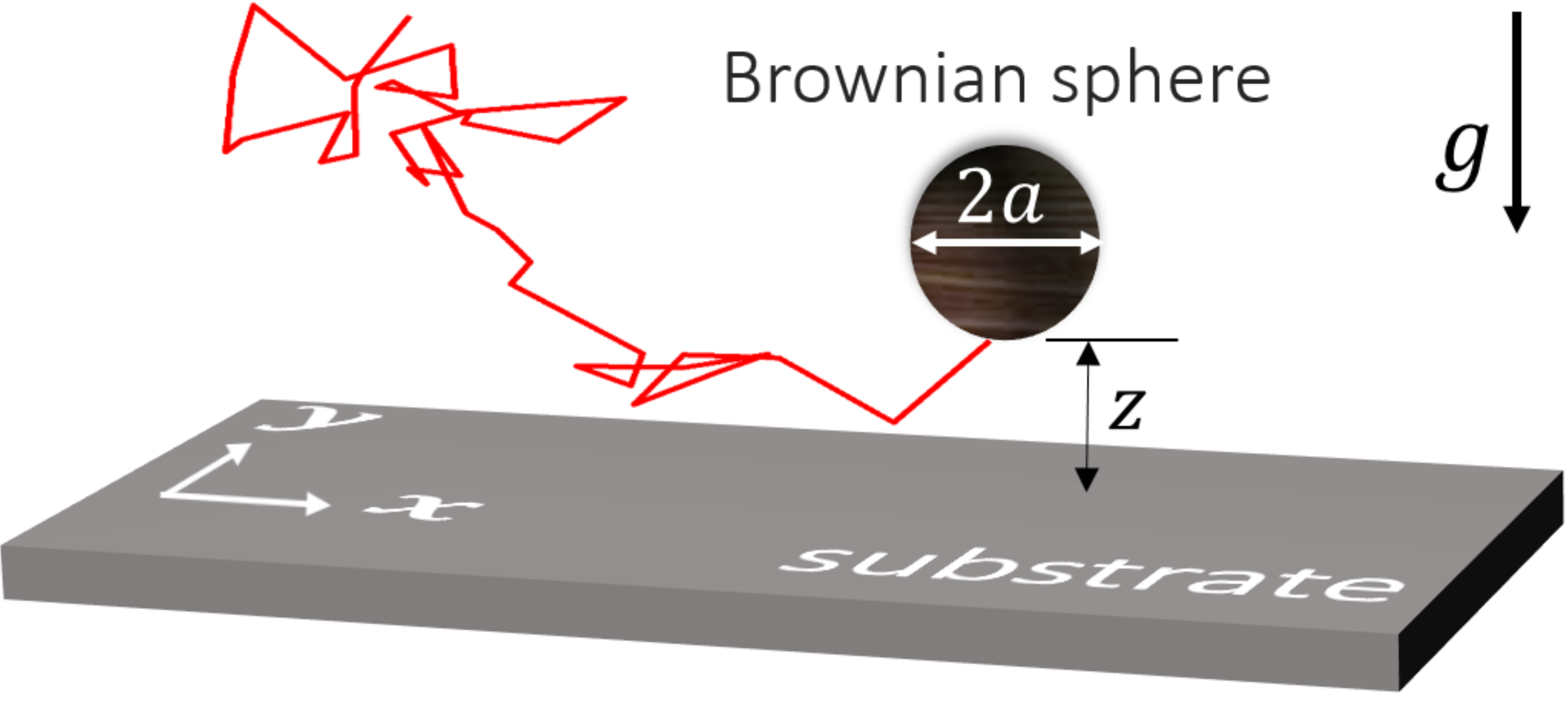}
	 \end{center}
	 \caption[example] {\label{wall} (color online). Brownian diffusion near a horizontal substrate.  }
\end{figure}

For our experiments, we consider the Brownian motion of a colloidal sphere near a planar horizontal surface (Fig.~\ref{wall}), a situation where the diffusion constant  varies in a known way with distance from the surface. For a freely diffusing Brownian sphere in an unbounded fluid medium, the diffusivity is given by the Stokes-Einstein relation,
$
D_0=\frac{k_BT}{6\pi\eta a},
$
where  $k_B$ is  Boltzmann's constant, $T$ the temperature, $a$  the particle radius,  and $\eta$ the fluid's dynamic viscosity.  For diffusion near a solid planar surface, theoretical \cite{Theory1,Theory2,Theory3} and experimental \cite{Exp0,Exp1,Exp2,Exp3,Exp4,Exp5,Exp6} studies have shown that the diffusivity decreases anisotropically with distance $z$ from the plane, owing to the hydrodynamic interaction between the sphere and the plane.  A useful second-order Pad\'e approximation \cite{Exp0} to the infinite-series results found by Brenner \cite{Theory2} gives the vertical diffusivity 
\begin{align}
	D_\perp(z) &\cong D_0 \left( \frac{6z^2+2az}{6z^2+9az+2a^2} \right)  \,,
\label{eq:Dperp}
\end{align}
where $z$ is the height of the bead bottom above the substrate (Fig.~\ref{wall}).  For small $z$, we have $D_\perp / D_0 \approx z/a$.

By contrast, for small $z$, the value of $D_\parallel$ is significantly higher and its relative variation much smaller (except perhaps extremely close to the plane \cite{Theory2}).  As we will see, the stronger relative variation of $D_\perp$ can lead to non-Gaussian dynamics, whereas the weaker relative variation in $D_\parallel$ does not generate measurable deviations from Gaussian displacement distributions \cite{Supp}. The experiments presented below consider vertical motion only.

Near a horizontal substrate, the vertical motion of a sphere is influenced by both  gravitational and electrostatic forces.  The surface of the colloidal particle and the substrate in a liquid may carry ionized chemical groups, which in our experiments lead to repulsive electrostatic double-layer forces that  prevent the Brownian particles from sticking to each other and to the surface of the substrate \cite{Forces}.  For large-enough heights, van der Waals forces can be neglected since they are very short ranged (order of a few nanometers) and are masked by the longer-ranged double-layer forces (50--100 nm). The total potential energy $U(z)$ of a diffusing particle is therefore dominated by the gravitational field at larger heights and the double-layer potential at smaller heights:
\begin{align}
	\frac{U(z)}{k_BT} = 
	\begin{cases}
		\bar{B} \, \mathrm{e}^{- {z}/{\ell_D}} + z / \ell_g \,, & z \ge 0 \,, \\
		\infty \,, & z<0 \,,
	\end{cases}
\label{uz}
\end{align}
where $\ell_D $ is the \textit{Debye  length}, which measures the effectiveness of the screening---the range of double-layer interaction effects.  In experiments, we chose a Debye length large enough to keep the sphere-substrate interactions simple by eliminating  van der Waals forces.   (A larger $\ell_D$ also keeps characteristic times longer, making dynamical behavior easier to measure \cite{Supp}.)  The prefactor $\bar{B}$ measures the strength of the double-layer potential, in units of $k_B T$.  Finally, the \textit{gravitational decay length} 
$
\ell_g = \frac{k_BT}{\Delta mg}
$
is the typical distance moved by the particle in the gravitational potential in response to thermal forces. Here, $\Delta m =\Delta \rho\left(\frac{4}{3}\pi a^3\right)$ is the mass difference between  particle and displaced solvent, and $\Delta \rho$ is the corresponding density difference.

Given the potential (\ref{uz}), the motion of the particle is described by the overdamped Langevin equation, according to which its displacement in a short interval $\delta t$ is given by \cite{drift2, Supp,lau07}
\begin{align}
\label{pm}
	\Delta z  \approx \left[ \frac{\mathrm{d}D_\perp(z)}{\mathrm{d}z} -\frac{D_\perp(z)}{k_BT}
		\frac{\mathrm{d} U(z)}{\mathrm{d} z} \right] \delta t +\sqrt{2D_\perp(z)\delta t}\; \xi \,,
\end{align}
where we use the \textit{isothermal rule} for stochastic integration \cite{lancon02, drift2}. Here, $\xi$ is a Gaussian random variable satisfying $\braket{\xi}=0$ and $\braket{\xi^2}=1$.

\begin{figure}
	 \begin{center}
	 \includegraphics[width=8.4cm]{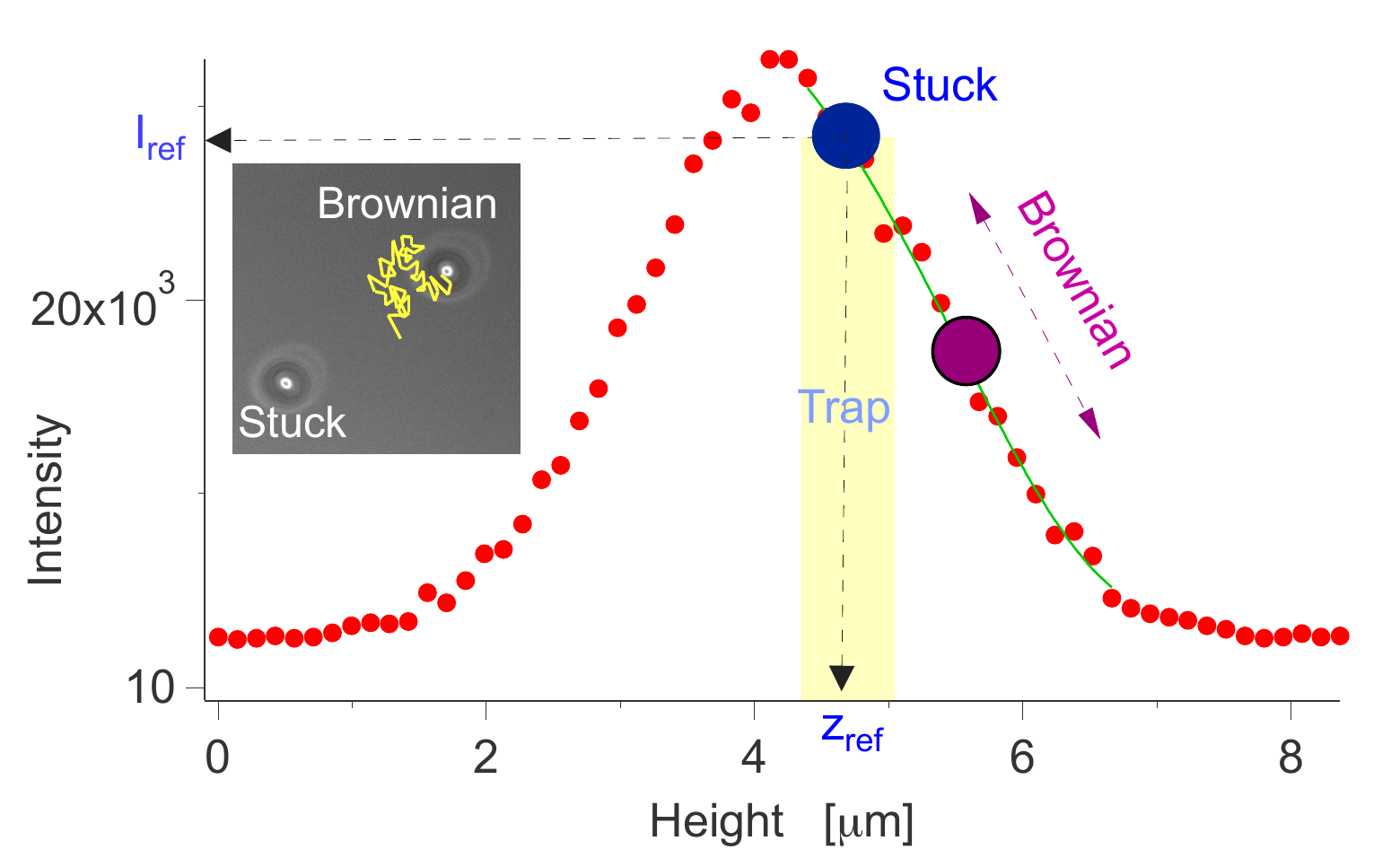}
	 \end{center}
	 \caption[example] { \label{profile} (color online). 
Intensity profile of the central region of a bead stuck to the microscope slide. The stuck bead was kept at height $z_{\rm ref}$ by a feedback loop, while the Brownian bead diffused above the stuck bead.  Inset shows camera image of both stuck and Brownian (freely diffusing) beads.}
\end{figure}

Apart from the above-mentioned three length parameters---$a$, $\ell_D$ and $\ell_g$---that can be controlled in experiments, the time interval $\Delta t$ over which the displacements are measured is important.  To observe the diffusing-diffusivity mechanism, we need to ensure that variations in $D$ are  the dominant contribution to non-Gaussian dynamics within  $\Delta t$. Since the bounding potential can lead to undesired nonlinear MSD, along with non-Gaussian dynamics, we work in a regime where thermal fluctuations  dominate over  deterministic drift.  For diffusion to dominate, we impose $\overline{\Delta t} \equiv \Delta t (D_0 / \ell_g^2) < \overline{\Delta t}_c$ \cite{Supp}.  A simple estimate based on a harmonic approximation to the potential of Eq.~\eqref{uz} then predicts $\overline{\Delta t}_c \approx 13$ for the parameters typical of our experiment \cite{Supp}.

Another requirement is that the experimental parameters should lead to non-Gaussian displacement distributions.  To measure deviations from a Gaussian distribution, we used the \textit{excess kurtosis} 
\begin{equation}
	\kappa \equiv \frac{\braket{\Delta {z}^4}}{[{\braket{\Delta {z}^2}}]^2} -3 \,,
\end{equation}
which is defined so that $\kappa=0$ for a Gaussian distribution and $\kappa>0$ for a heavier-tailed distribution \cite{kurt}.  We found that latex beads of radius 2.5~$\mu$m in purified water above a glass substrate gave an easily measurable diffusing-diffusivity effect for vertical motion. 

In our experimental setup, a bright-field microscope was used to image beads in three dimensions in reflection (details of the experimental design  in \cite{Supp}).  The bead's vertical position was inferred directly from  intensity images by averaging pixel intensities over a 3 $\times$ 3 pixel region centered on  the pixel with the maximum intensity. The bead's intensity profile  against height relative to the objective (Fig.~\ref{profile}) was obtained by vertically moving an immobilized bead, using a voltage applied to a calibrated piezo-stage and measuring the intensity of the bead from the images taken.  An intensity calibration was  performed before and after each set of measurements.

\begin{figure}
	 \begin{center}
	 \includegraphics[width=8.6cm]{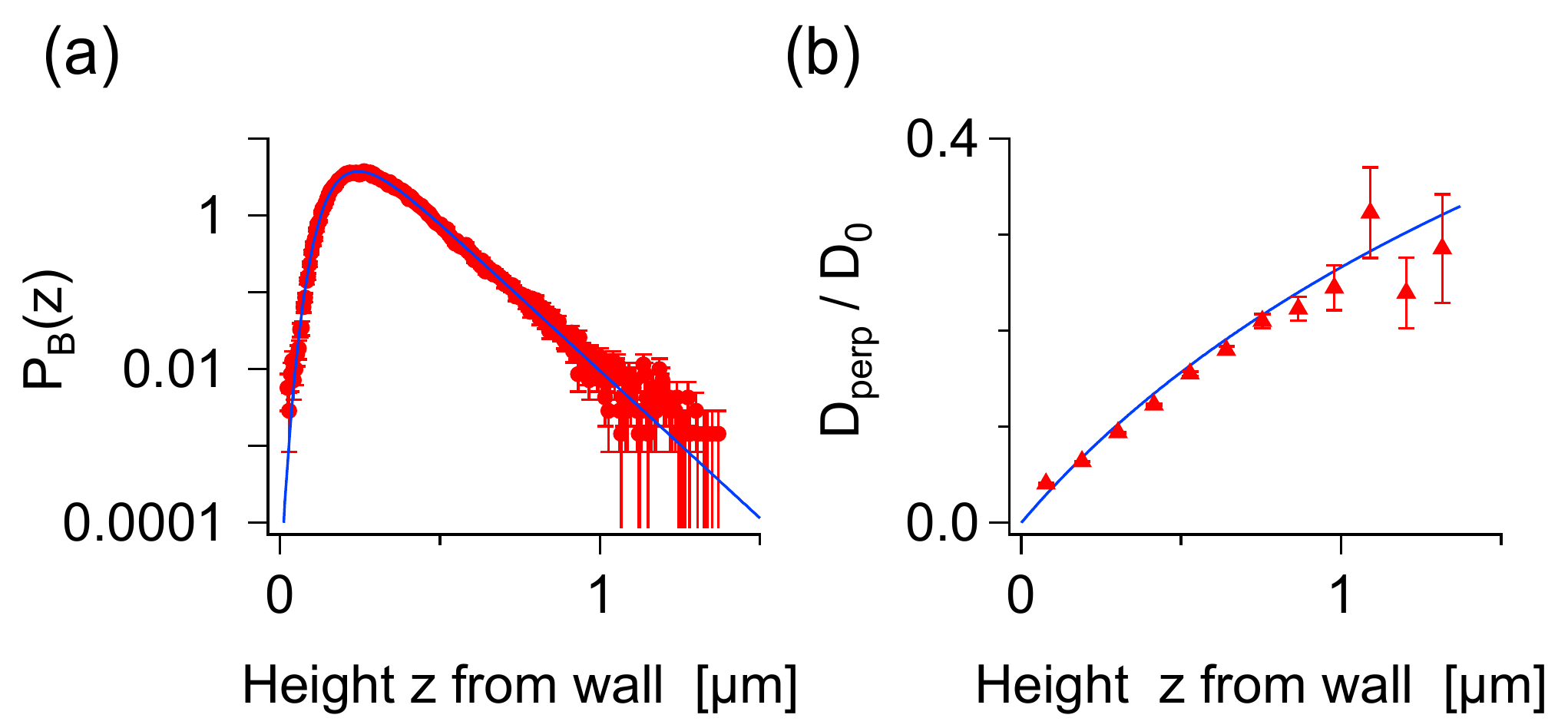}
	 \end{center}
	 \caption[example] { \label{statics} (color online). (a) Probability density function $P_B(z)$ of bead height above substrate, $z$.  Solid line is a fit to the Boltzmann distribution, Eq.~\eqref{uz}.    (b) Vertical diffusivity of the bead vs. $z$.  Solid line is a plot---not fit---of Eq.~\eqref{eq:Dperp}, using material parameters inferred from part (a).  Data from Run 10, with parameters from Table I \cite{Supp}.}
\end{figure}

To eliminate mechanical drift due to thermal expansion or contraction between the sample cell and the microscope objective, we also tracked an immobilized bead (stuck to the substrate) as a reference from which the height $z$ of the diffusing bead above the substrate could be determined using a fit of the intensity profile.  We set the objective-stage separation to be just out of focus, so that the intensity of the two beads was located on the right side of the profile (Fig.~\ref{profile}). This ensured that none of the beads had positions on the other side of the profile, which would lead to ambiguity in the bead's position, as the profile is almost symmetric at the peak.  We then applied a feedback loop to stabilize the position of the stuck bead.  By choosing larger beads ($r \approx 2.5~\mu$m), we ensured that the maximum height of the diffusing bead was always within the linear regime depicted by the thin green line in Fig.~\ref{profile}.  The differential measurement also eliminates noise due to variation of the light source.

The distribution of heights for one bead trajectory is shown in Fig.~\ref{statics}(a). The solid line is the least-squares fit to the Boltzmann distribution   $P_B(z)~\propto~\exp \left[{-U(z)/k_BT}\right]$, using Eq.~\eqref{uz} (the details are described in Ref.~\cite{Supp}).  The fit parameters were $\ell_g=0.113 \pm 0.005$ $\mu$m (expected value using nominal bead size from manufacturer $\approx 0.11 \pm 0.02$~$\mu$m), $\ell_D=0.079\pm 0.004$~$\mu$m, and $\bar{B} = 15.3 \pm 0.8$ \cite{Supp}.  The height reference $z_{\rm ref}$ (substrate position) is also fit.  Since $\ell_g$ is well determined from the data, we used it to infer the bead size.  For the data in Fig.~\ref{statics}(a), we find $a = 2.53 \pm 0.04$ $\mu$m.  The variation of the bead's vertical diffusivity with  height from the substrate  is shown in Fig.~\ref{statics}(b). The diffusion coefficients are measured using conditional displacements $\Delta z$ in a narrow interval (0.1 $\mu$m) centered on height $z$ above the substrate, correcting for the camera exposure $t_{\rm exp}$ \cite{Supp}.  Using the asymptotic diffusivity $D_0 = 0.0996 \pm 0.0015$ $\mu$m$^2/$s calculated from the inferred parameters, we plot the normalized diffusivity $D_\perp(z) / D_0$ from Eq.~\eqref{eq:Dperp} in Fig.~\ref{statics}(b), with no adjustable parameters.  

\begin{figure}
	 \begin{center}
	 \includegraphics[width=8.6cm]{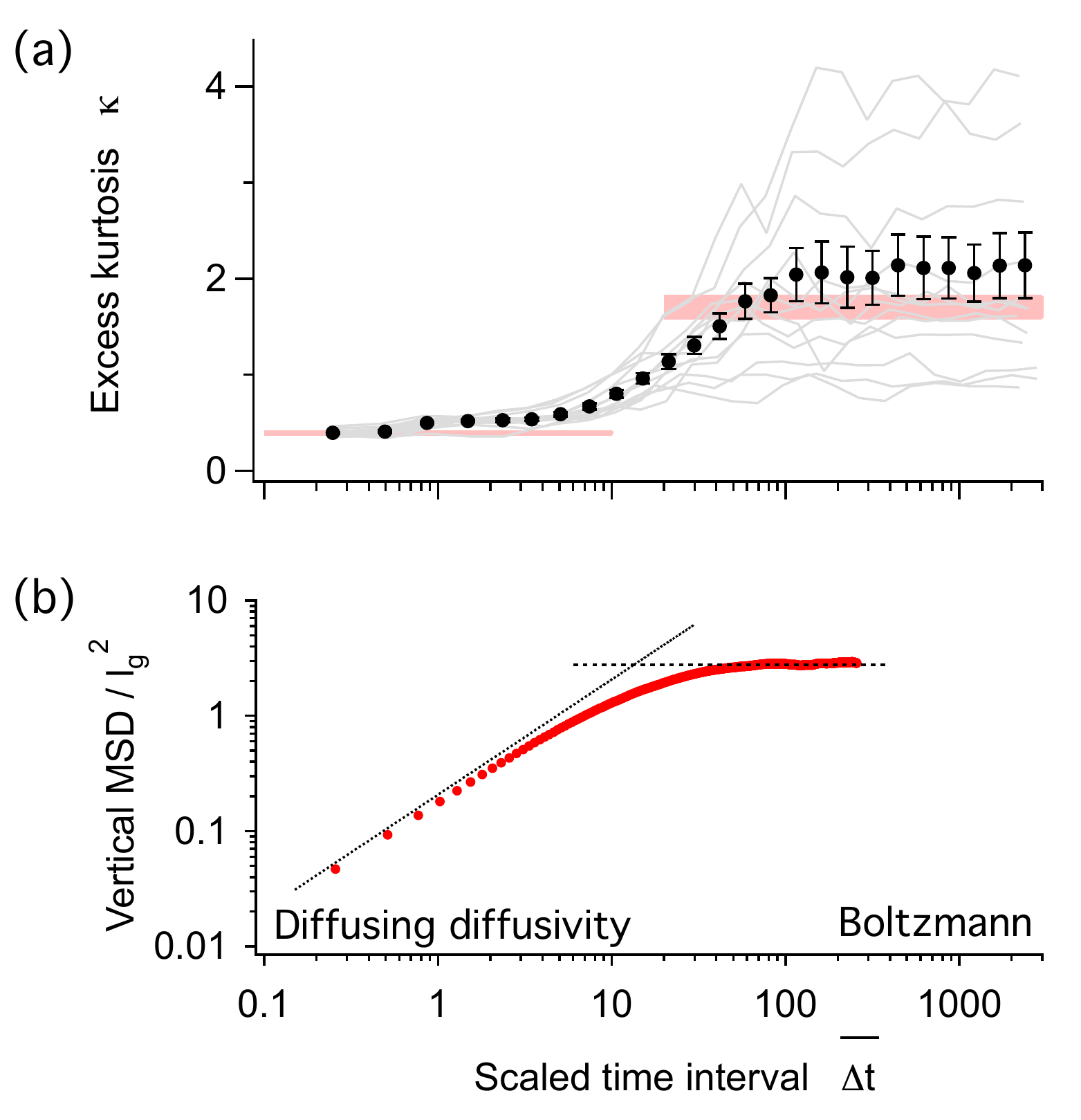}
	 \end{center}
	 \caption[example] { \label{kurtosis} (color online). Diffusing diffusivity and Boltzmann regimes for short and long time intervals, with $\overline{\Delta t} = \Delta t (D_0 / \ell_g^2)$.  (a) Excess kurtosis is non-zero for all intervals.  Light gray curves show results from 13 individual runs \cite{Supp}.  Solid symbols represent the unbiased estimate based on the 13 runs, with the standard error of this estimate shown, too.  Uncertainties in the scaled time interval $\overline{\Delta t}$ are $\approx 2\%$, which is smaller than the symbol size. Pink bars represent the diffusing-diffusivity limit at short time intervals and the Boltzmann limit at long time intervals.   (b) Mean-square displacement (MSD).   Data from Run 10.  Dashed line shows linear MSD behavior for short time intervals, with the slope obtained theoretically using parameters from Table I of~\cite{Supp}.  Dotted line shows MSD for the difference between two position measurements drawn from the Boltzmann distribution for $U(z)$, likewise calculated theoretically.}
\end{figure}

The excess kurtosis and MSD  at different time scales are shown in Fig.~\ref{kurtosis}.  For the excess kurtosis, the results of 13 different runs are in light gray, and the unbiased estimate obtained from them as described in Supplementary Material~\cite{Supp} is given by the solid symbols. To reduce statistical noise in the kurtosis, the data are binned, with the bin widths roughly constant on the logarithmic scale and data within each bin are averaged (or, rather, the same unbiased estimate procedure is applied) and assigned to the value of $\overline{\Delta t}$ corresponding to the average between the values within the bin. The displacements are non-Gaussian at all time scales, with heavier, nearly exponential tails, as illustrated by the histogram at left in Fig.~\ref{heights} and also in the Supplemental Material \cite{Supp}.  The excess kurtosis interpolates between $\approx 1.9$ at large time intervals $\overline{\Delta t}$ to $\approx 0.4$ at small time intervals, with a crossover at $\overline{\Delta t}_c \approx 20$, which is close to the value calculated theoretically~\cite{Supp}.  We notice that for each run the fluctuations for different values of $\overline{\Delta t}$ are correlated, because each point on a single curve is calculated from the same time series.  

The two observed plateaus in kurtosis suggest that there are two regimes:  diffusing-diffusivity ($\Delta t<\Delta t_c$) and Boltzmann ($\Delta t>\Delta t_c$).  In the diffusing-diffusivity regime, the non-Gaussian dynamics is driven by $D$ variations, and the bounding potential has negligible influence.  In this regime, the displacements are non-Gaussian; yet the MSD grows linearly with time [Fig.~\ref{kurtosis}(b)].  The non-Gaussian displacement distribution is generated by the diffusivity distribution $P(D)$.  Then, for $\Delta t \to 0$,  
\begin{equation}
\label{PD}
	P(\Delta z ; \Delta t) \approx \int_0^\infty \mathrm{d}D \, P(D) \, \frac{1}{\sqrt{4\pi D\Delta t}}
	\exp \left[-	\frac{\Delta z^2}{4D\Delta t}\right] \,.
\end{equation}
In Fig.~\ref{kurtosis}(a), we see that the kurtosis is constant near the minimum time scales probed in the experiment.  At these short time scales, $P(\Delta z)$ is governed chiefly by the time-independent $P(D)$.  If $D$ did not vary, we would expect no kurtosis (as observed for horizontal displacements \cite{Supp}).

In the Boltzmann regime, the bounding potential dominates the $D$ variations, and the MSD saturates [Fig.~\ref{kurtosis}(b)].  At very large time intervals, we can view each position measurement as an independent sample from the equilibrium Boltzmann distribution.
For $\Delta t \to \infty$,
\begin{equation}
\label{Pz}
	P(\Delta z) \approx \int_{-\infty}^\infty \mathrm{d}z \, P_B(z) \, P_B(z+\Delta z) \,,
\end{equation}
where the potential $U(z)$ in Eq.~\eqref{uz} and hence the Boltzmann distribution $P_B(z)$ is formally defined for all $z$, with $P_B(z)=0$ for $z<0$.  

For the bounding potential in the system we study,  Eq.~\eqref{Pz} is clearly non-Gaussian (it has exponential tails $\propto {\rm e}^{-|\Delta z|/\ell_g}$).  The predicted values of the excess kurtosis for the 13 experimental runs based on the parameter estimates for these runs (see Table I in Ref.~\cite{Supp}) range from 1.42 to 1.82, with the average 1.70.  The range is consistent with the unbiased estimate at large time intervals based on the 13 runs (Fig.~\ref{kurtosis}), keeping in mind the small number of runs and the approximations made when estimating the uncertainties.  The large spread between the runs for large $\Delta t$ is mostly due to lack of statistics and is consistent with the results of numerical simulations \cite{Supp}.

Finally, we divided the vertical-position measurements of the bead into very small height intervals ($\approx 0.01\;\mu$m) in the diffusing-diffusivity regime ($\Delta t = 0.033$ s) and studied the displacement distribution in each interval (Fig.~\ref{heights}).  The results directly confirm the diffusing-diffusivity mechanism predicted by Eq.~\eqref{PD}:  The displacements are nearly Gaussian at each interval with different variances (right side of Fig.~\ref{heights}); yet the overall distribution is non-Gaussian (left side of Fig.~\ref{heights}).

\begin{figure}
	 \begin{center}
	 \includegraphics[width=8.6cm]{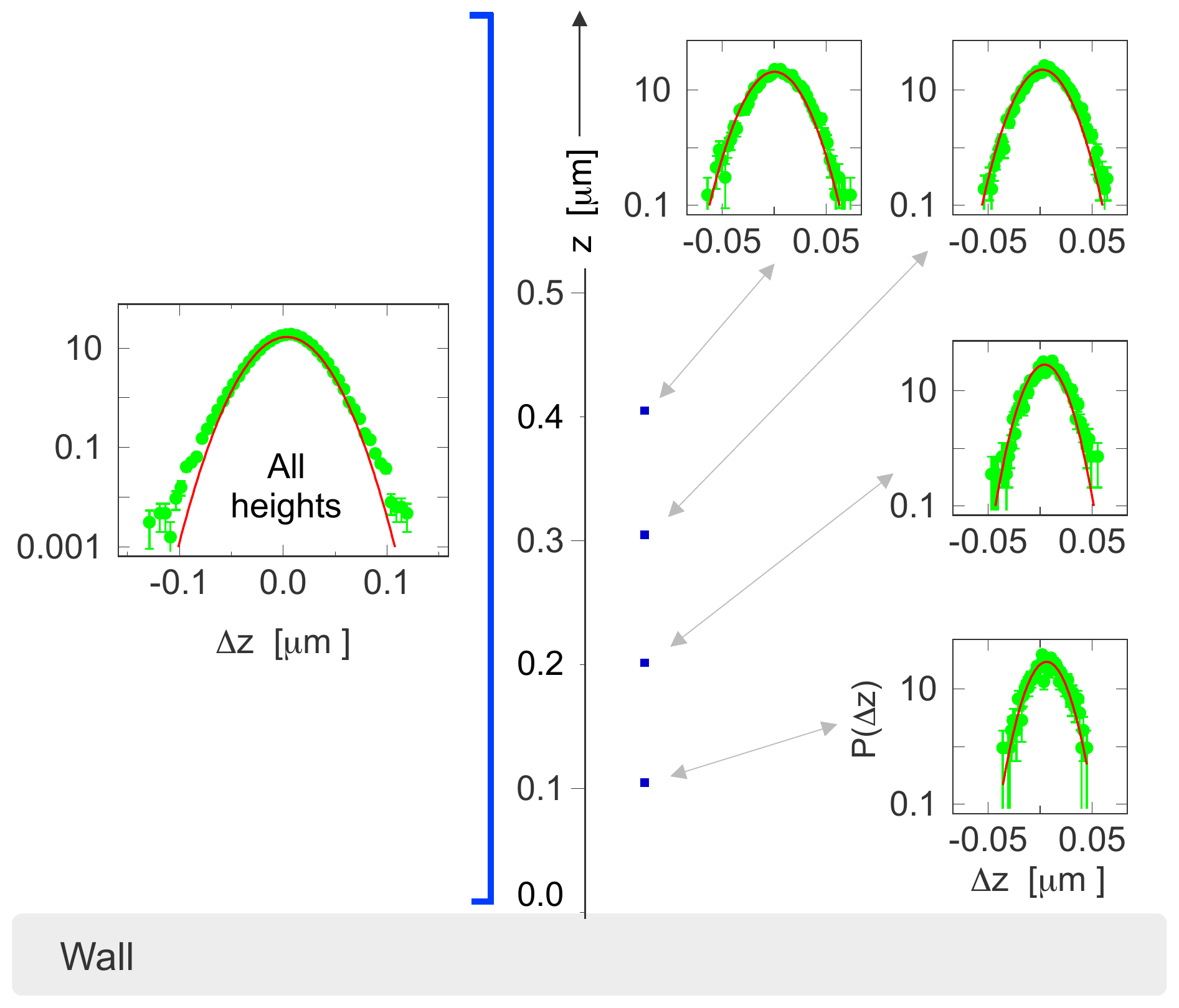}
	 \end{center}
	 \caption[example] {(color online). Displacements at starting-point intervals of $\approx 0.01\;\mu$m close to the substrate. Right side shows the nearly Gaussian  height conditional distributions.  Left side shows the non-Gaussian distribution from all heights explored by the bead.  $\Delta t = 0.033$ s.  Data from Run 10 \cite{Supp}.}
\label{heights}
\end{figure}

In conclusion, we have investigated experimentally the Brownian motion of colloidal spheres near a planar surface and have made the first direct confirmation of the diffusing-diffusivity mechanism \cite{NG12}:  At small time intervals, non-Gaussian displacements coexist with a MSD that grows linearly with time.  Our experimental system is unique among studies of this mechanism in that we  independently measure the local value of the diffusivity, thereby showing that the conditional distributions at small height intervals from the substrate exhibit nearly Gaussian displacement distributions. 


The results of the experiments reported here give a rigorous test of the diffusing-diffusivity mechanism in a model system with ``quenched disorder'' where diffusivity variations are measured independently.  They suggest new ways to understand the behavior of other systems near walls and interfaces, such as the collective motion of sperm cells near interfaces \cite{app1} and swimming bacteria in thin films \cite{app2}.  These quantitative results from a simple model system also give confidence in the more qualitative analyses done on more complex systems \cite{NG1, NG2, NG3,  NG5, NG6, NG7, NG8, NG9, NG10} where the same phenomena---linear MSD and non-Gaussian displacements---are observed.  


We thank Gary  Slater and Maxime Ignacio for helpful discussions.  We also thank Paul Omelchenko, Lukas Schertel, and Dirk Wiedmann for their contributions to the experimental apparatus. This work was supported by NSERC (Canada).  MVC also acknowledges support by Leverhulme Trust through a grant to J. E. Sprittles.


\clearpage

\begin{center}
\textbf{\large Supplemental Materials:  Test of the diffusing-diffusivity mechanism using near-wall colloidal dynamics}
\end{center}
\setcounter{equation}{0}
\setcounter{figure}{0}
\setcounter{table}{0}
\setcounter{page}{1}
\makeatletter
\renewcommand{\theequation}{S\arabic{equation}}
\renewcommand{\thefigure}{S\arabic{figure}}
\renewcommand{\bibnumfmt}[1]{[S#1]}
\renewcommand{\citenumfont}[1]{S#1}

\section{Experimental Setup}
\begin{figure}
	 \begin{center}
	 \includegraphics[width=3.5in]{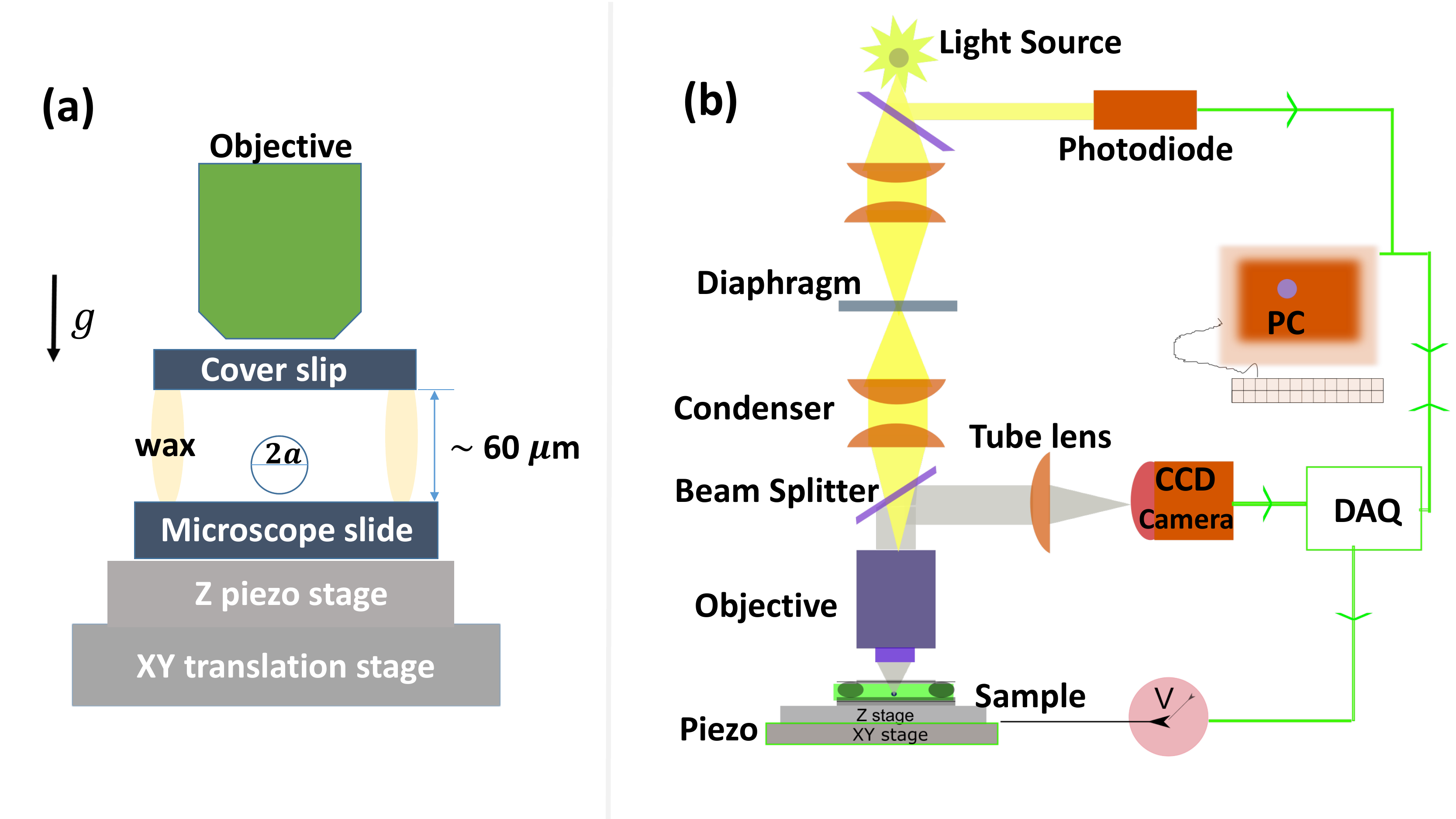}
	 \end{center}
	 \caption[example] {\label{setup}(color online). Schematic diagram of experimental setup. (a). Details of sample. (b) Microscope and feedback loop used to track the beads.  }
\end{figure}

We assembled a vertically aligned, bright-field microscope that imaged in  reflection (Fig.~\ref{setup}). A halogen bulb fiber-optic illumination (Model 190 Fiber-Lite  Halogen Illuminator, Dela-Jenner Industries, Boxborough  MA,  U.S.A.) was used as a light source, made more uniform by  putting a ground-glass plate in front of it.  We used  a 60X  water-immersion objective (UPlanSApo, NA=1.2, Olympus Corporation, Shinjuku, Tokyo,  Japan). 

Samples were prepared as follows: Latex spheres were diluted and mixed with purified water, at  volume fractions   low enough that each bead can be considered to move independently from all others.   Sample chambers ($\approx$ 60--80 $\mu$m in thickness) were made  by placing four pieces of Parafilm  (Bemis Co., Neenah WI, U.S.A.) between a microscope slide (1 mm thick) and a No.~1 coverslip ($\approx$ 0.17 mm thick).   The coverslip was first cleaned using a nitrogen gas ionizing gun  (Top Gun Static Neutralizer, SIMCO Inc., Hatfield PA, U.S.A.) before use.  The cell was partially sealed using  Parafilm  melted on a hot plate, then filled (without bubbles) with the beads in solution. Finally, the cell was completely sealed with melted wax in order  to avoid fluid flow due to evaporation or convection and allowed to cool.  The sample was then placed on an XY translation stage  (Model 406,   Newport Corporation,  MT, Irvine CA, U.S.A.), which  moved the sample to search for   beads. A feedback-controlled piezo stage  (Nano OP-65, controlled by Nano-drive 85, Mad City Labs, Madison WI, U.S.A.) was used for tracking the $z$-positions of the bead.

Images were recorded by a CCD camera  (Model FL3-FW-03S1M-C, {Point Grey Research}, Richmond BC,  Canada). The image acquired was processed via the computer, using  {LabVIEW} software (National Instruments, Austin TX, U.S.A.), to determine the position of the bead. The camera was triggered by the rising edge of a square wave from a function generator  (2 MHz Function Generator, Model 3011b, B$\&$K Precision Corporation, Yorba Linda CA, U.S.A.).  The camera's frame rate was set to  30 Hz and the shutter speed to $t_{\rm exp} =$ 10 ms.  Intensities were digitized at 12-bit resolution ($2^{12}=4096)$ and mapped onto a 16-bit intensity scale ($2^{16}=65536$). The field of view of the camera was about 60~$\mu$m $\times$ 40~$\mu$m. 

We acquired and processed images at fixed time intervals (33~ms) to determine the bead's trajectory.  If the monitored bead   diffused out of the field of view, we used the XY stage to manually reset the bead's image to the center of the field of view and continued collecting data on the same bead. In practice, we could obtain trajectories up to about 10$^4$ s. 

To track the position of the bead in the XY-plane, we used the edge-detecting algorithm from the IMAQ Vision module for LabVIEW.  The vertical position was obtained from intensity variations, as mentioned in the main paper.  The feedback loop used for estimating the vertical positions follows these steps: For each image taken, the LabVIEW program determines the stuck bead's intensity $I$.  The slope $b$ from the linear fit is used to calculate a required voltage $V=\alpha(I-I_{\rm ref})/b$, where $\alpha$ is the feedback gain,  to move the stuck bead back to the set plane (at $I_{\rm ref}$).  Using a National Instruments data acquisition device (NI-USB 6215), the voltage is applied, and the iteration continues.  

To correct for uneven illumination, which could bias our height measurements of the Brownian bead, we normalized the bead intensity pixel by pixel, using the intensity of an averaged background image.  This procedure was done before and repeated after the Brownian bead measurements were taken, to ensure that the background image stayed uniform throughout the experiment.  The bead's intensity values were further normalized by the illumination intensity recorded by a photodiode.  We eliminated the effects of stray ambient light by covering the microscope with a box. 

In order to estimate the variation of the diffusion constant with the height $z$ above the substrate, we measured displacements $\Delta x$ and $\Delta z$ conditioned on starting in a given height interval on bin $n$ of width $\Delta z_{\rm bin}$.  That is, $z \in (z_{{\rm bin}~n}, z_{{\rm bin}~n} + \Delta z_{\rm bin})$.  Denoting the mean-square conditional vertical displacement by
\begin{align}
	\left< (\Delta z)^2 \right>_n \,,
\end{align}
The estimate of $D_{\rm perp}(z)$ for bin $n$ is
\begin{align}
	D_{\rm perp}(z)_n 
		= \frac{\left< (\Delta z)^2 \right>_n - 2 \xi^2)}{2 \left( \Delta t - \frac{1}{3}t_{\rm exp} \right)} \,,
\label{eq:Dperp_est}
\end{align}
where the denominator corrects for the blurring effects of the camera exposure, $t_{\rm exp} = 10$ ms, which is not small compared to the measurement interval $\Delta t = 33$ ms.  The $\xi^2$ term represents the variance of the measurement noise (assumed uncorrelated and identically distributed at different times), which is estimated $\xi \approx 0.004~\mu$m by extrapolating the autocorrelation function of position measurements as a function of shift and isolating the ``extra" variance at zero shift.  See \cite{goulian00,savin05,cohenThesis} for derivation and discussion of the motion-blur effect.  Since the measurement noise in the time series for $z$ was quite small, it had negligible effect on the diffusion-constant estimates.

\section{Supplemental Experimental Results:  Vertical displacements}

\begin{figure}[h]
	 \begin{center}
	 \includegraphics[width=3.5in]{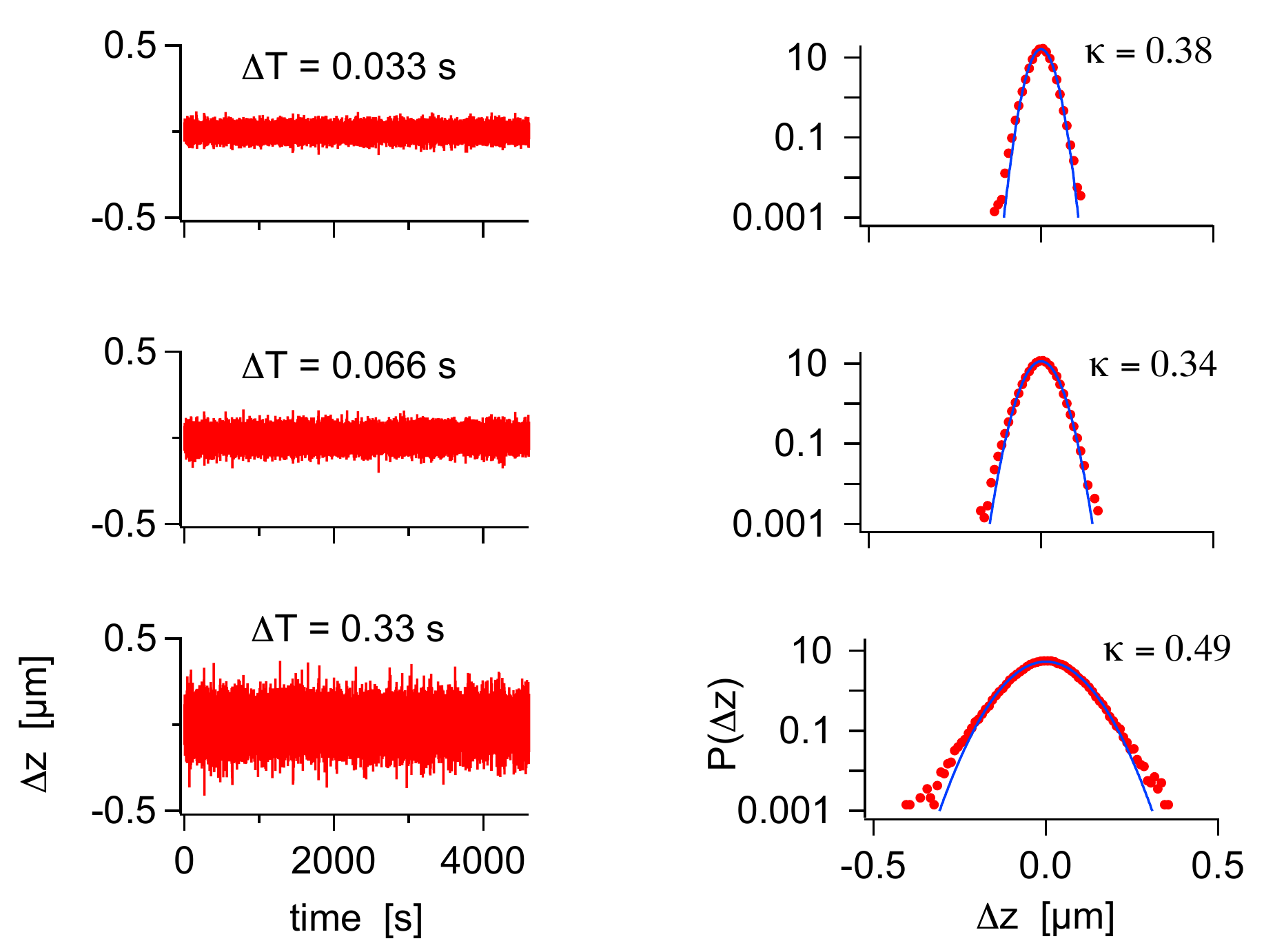}
	 \end{center}
	 \caption{(color online). Time series of vertical displacements and histograms.  Data from Run 10.}
\label{fig:vert}
\end{figure}

Figure~\ref{fig:vert} shows time series and vertical displacements for three different time intervals.  The slight deviation in the tails from a Gaussian distribution leads to a positive excess kurtosis.

For Fig.~4 in the main text, we collected 13 sets of data for different trajectories averaging about 45 minutes (82 000 data points) each.   The results are summarized in Table \ref{ptable2}.  

\begin{table}
\begin{center}
\begin{tabular}{c|r|c|c|c|c|c }

Run& $N \quad$ & $l_g$  [$\mu$m]  &  $l_D$  [$\mu$m] & $\, \frac{B}{k_BT}$&
	$a$ [$\mu$m]&$D_0$ [$\mu$m$^2$/s]  \\ \hline

1&    68 966                   &0.120                    &0.071                           &16.8                  &2.48  &  0.102                      \\

2&    71 628                  &0.118                      &0.077                           &15.6                    &2.49       &  0.101                                        \\

3&    100 394                 &0.117                     &0.071                           &17.7                    &2.50          &  0.101                                     \\

4&    119 155                  &0.121                      &0.078                           &14.7                    &2.47          & 0.102                            \\

5&    93 778                  &0.117                       &0.075                           &16.3                    &2.50          &  0.101                                  \\

6&    62 869                  &0.114                      &0.070                           &17.0                    &2.52      &  0.100                                   \\

7&    38 763                  &0.121                      &0.073                           &15.8                    &2.47      &   0.102                             \\

8&    34 892                  &0.105                       &0.086                           &14.5                    &2.59       &   0.097                              \\

9&    80 065                  &0.117                       &0.074                           &16.2                    &2.50      &  0.101                                \\

10&   139 716                   &0.113                       &0.079                           &15.3                    &2.53    &    0.100                              \\

11&   119 645                   &0.123                       &0.072                           &16.8                    &2.46    &   0.102                                 \\

12&   82 324                   &0.118                       &0.072                          &15.6                    &2.49     &   0.101                            \\

13&   54 923                   &0.106                       &0.079                           &16.4                    &2.58     &   0.098                           \\

\end{tabular}
\end{center}

 \caption[Experimentally determined parameters from 13 runs]{Values obtained for experimental parameters from 13 different runs. $N$ gives the number of data points in the run.  The radius $a$ is inferred from $\ell_g$ using the measured temperature $T = (298.5 \pm 1) K$, and latex bead density $\rho_\text{b}= 1.055$ g/cc. The $D_0$ values also use the water viscosity $\eta = (0.88 \pm 0.02) \times 10^{-3}$ N s/m$^2$.}
\label{ptable2}
\end{table}

A histogram estimate of the position probability density for each run was fit separately to the Boltzmann distribution to obtain 13 sets of the parameters $\ell_g$, $\ell_D$ and $B$~\cite{uncert}; the values of $a$ and $D_0$ are then obtained from $\ell_g$.  For these histograms, a bin width of 5~nm was chosen, which is sufficiently small that the probability density does not change significantly over that length.  A weighted least-squares fitting procedure was used, with the weights inversely proportional to the number of data points in the bin and the bins with zero points ignored.  Estimating the uncertainties of the parameters is non-trivial because of correlations between the bins;  using simulations, as described in the corresponding section below, we estimate the uncertainties of $\ell_g$, $\ell_D$ and $\bar{B}$ as 5~nm, 4~nm and 0.8, respectively, for the longest runs and 10~nm, 8~nm and 1.7 for the shortest runs. The corresponding uncertainty of $a$ is 35--70 nm, while for $D_0$ it is 0.0015--0.003 $\mu$m$^2$/s.

The resulting sets of parameters are given in Table~\ref{ptable2}.  We see that the values are consistent with expected bead-to-bead variations and with theory expectations.  Notice that the variations of the parameters between the runs are somewhat larger than the statistical uncertainties of individual measurements given above.  The larger variation suggests that it arises mainly from bead-to-bead differences.  The amount of variation seen in the radius ($\approx \pm 2\%$) inferred in different runs is typical of manufacturer specifications for the coefficient of variation of the diameter.

Finally, we mention an important point in our analysis. Consider an imaginary series of runs using the same particle under the same conditions. We will refer to the run-to-run variations of estimates of various quantities in such a series as their statistical fluctuations. For quantities such as the bead radius $a$ or asymptotic diffusion constant $D_0$ these statistical fluctuations are smaller than the bead-to-bead variations of the true values of these quantities in the 13 actual runs we analyze. However, they are \textit{larger} in the case of the excess kurtosis $\kappa$, where large statistical fluctuations arise because $\kappa$ depends on the fourth moment of the probability distribution. Consequently, it makes sense to report a single estimate of $\kappa$ based on the 13 runs, as we do in the main text. A proper way to obtain this estimate is described below. But for the mean-square displacement (MSD), which depends on only the second moment, and for the relation between diffusivity and height, one should look at the data from a single run.


\section{Supplemental Experimental Results: Horizontal displacements}

\begin{figure}[h]
	 \begin{center}
	 \includegraphics[width=2.5in]{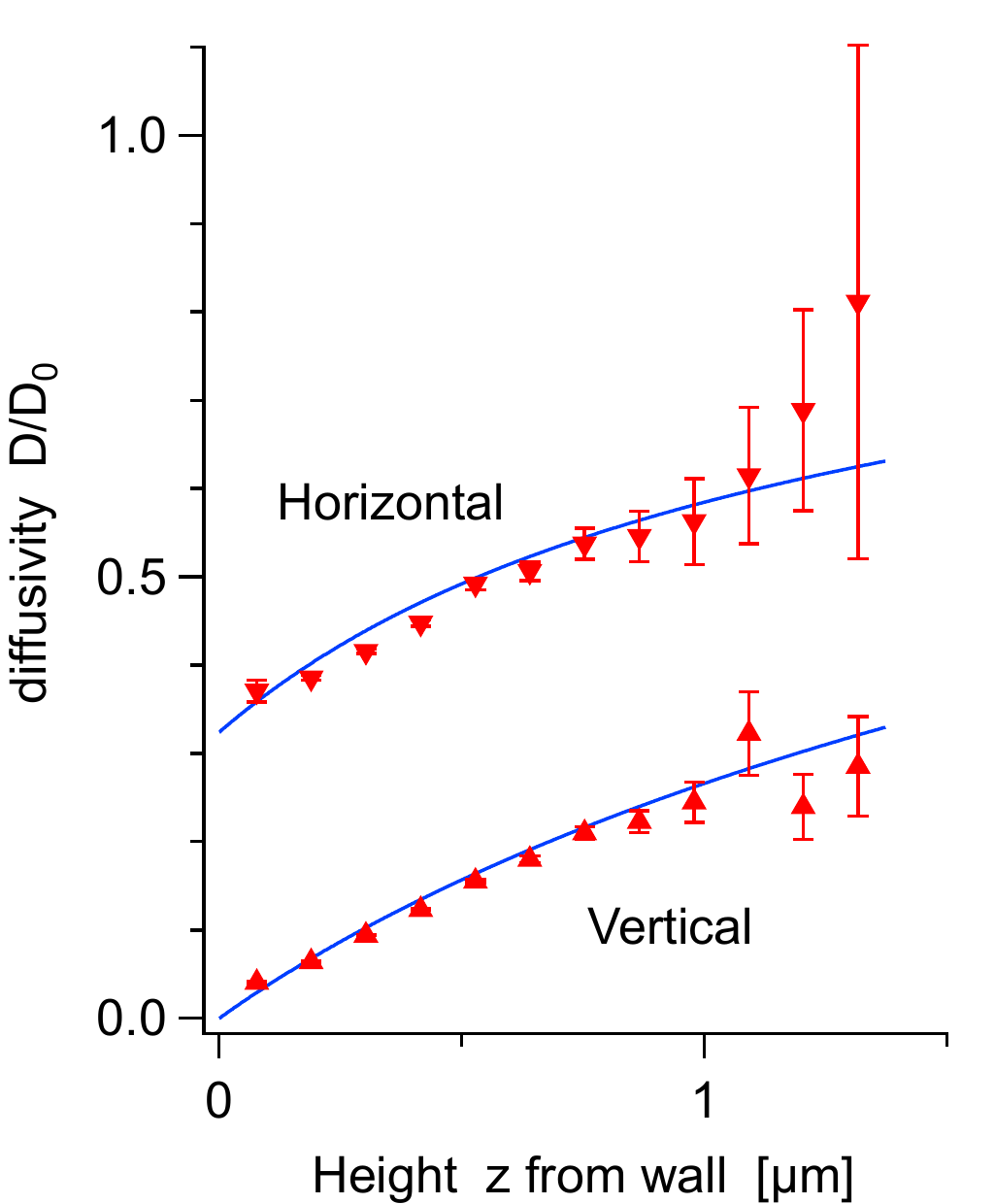}
	 \end{center}
	 \caption{(color online). Scaled horizontal and vertical diffusion coefficients, as a function of height $z$.  Solid lines are plots---not fits---of Eqs. (1) and \eqref{dpar}, based on parameters from Table~\ref{ptable2}.  Data from Run 10.}
\label{fig:heightStatshor}
\end{figure}

We first attempted to study the bead's diffusing-diffusivity mechanism using horizontal displacements. Since there is no confining potential for horizontal motion, the diffusing-diffusivity mechanism does not compete with the effects linked to the potential.  Figure~\ref{fig:heightStatshor} compares the measured dimensionless horizontal and vertical diffusion coefficients, along with analytical estimates.  The latter are given by Eq.~1 in the main text and by \cite{Theory2}.  For sphere radius $a$ and height above the substrate $z$, we define $\delta \equiv \tfrac{a}{a+z}$.  Then,

\begin{equation}
	\frac{D_\parallel(z)}{D_0} = 1 - \frac{9}{16} \delta + \frac{1}{8} \delta^3 
		- \frac{45}{256} \delta^4 - \frac{1}{16} \delta^5 + \mathcal{O} \left( \delta^6 \right) \,,
\label{dpar}
\end{equation}
for motion parallel to the substrate (e.g., along the $x$ direction).  As expected, $D_\parallel(z)$ has a higher value and smaller relative variation in the range of heights that we explore experimentally.  As a result, $D$ fluctuations are smaller, as is the diffusing-diffusivity effect.

\begin{figure}[h]
	 \begin{center}
	 \includegraphics[width=3.5in]{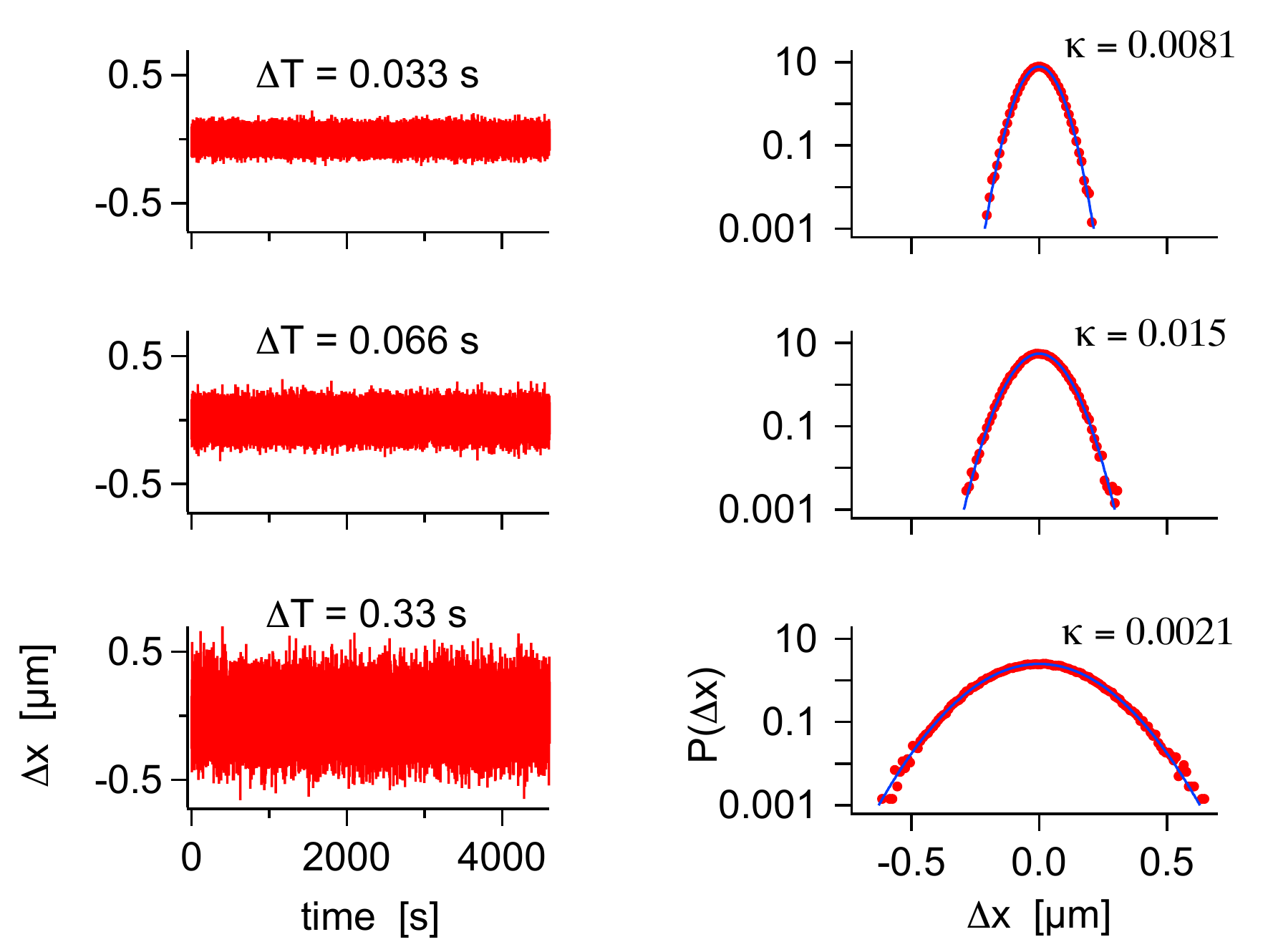}
	 \end{center}
	 \caption{(color online). Time series of horizontal displacements and histograms.  Data from Run 10.}
\label{fig:hor}
\end{figure}

\begin{figure}[h]
	 \begin{center}
	 \includegraphics[width=3.5in]{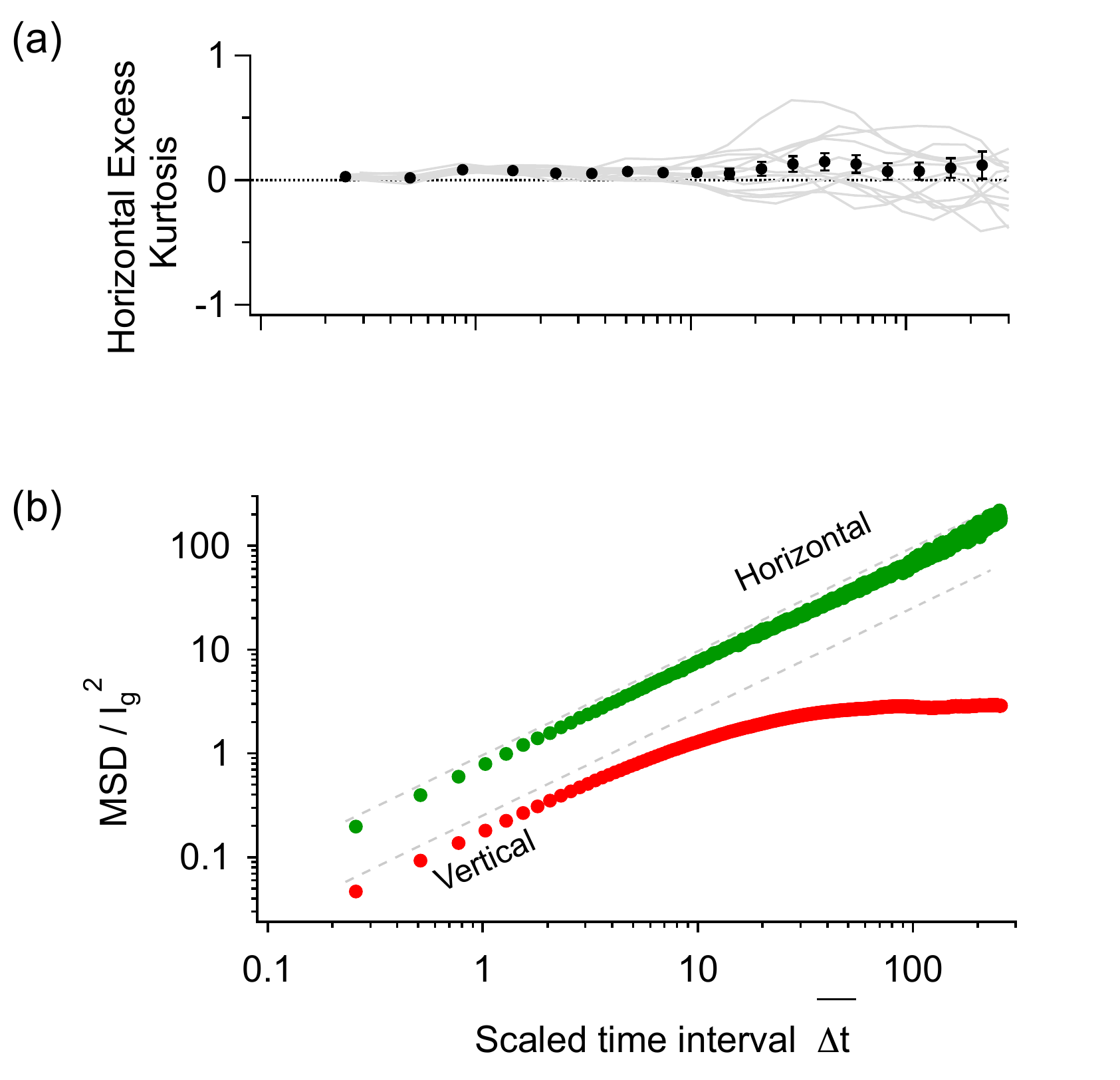}
	 \end{center}
	 \caption[example] {(color online). Results for horizontal diffusion: (a) Excess kurtosis: individual runs (gray curves) and unbiased estimates (solid symbols with error bars).  (b)  MSD as a function of scaled time intervals $\overline{\Delta t} = \Delta t (D_0/\ell_g^2)$ for horizontal and vertical motions.  Data from Run 10.}
\label{kurtosisMSDhor}
\end{figure}

As we see in Fig.~\ref{fig:hor} and more systematically in Fig.~\ref{kurtosisMSDhor}, horizontal  displacement distributions are nearly Gaussian.  The excess kurtosis [Fig.~\ref{kurtosisMSDhor}(a)] is nearly zero for all times. The MSD remains linear with time [Fig.~\ref{kurtosisMSDhor}(b)] for all explored time intervals.  Any trace of the diffusing-diffusivity mechanism is, unfortunately, too small to measure.

For more discussion of these results, see \cite{matse}.


\section{An unbiased estimate of the excess kurtosis}

The $i$th moment of a random quantity $\xi$ is defined as
\begin{equation}
m_i=\av{\xi^i}.
\end{equation}
In practice, moments are estimated by taking a finite number $N$ of samples from the distribution:
\begin{equation}
\widetilde{m_i}=\frac{1}{N}\sum_{j=1}^N \xi_j^i.\label{momest}
\end{equation}
The average of this estimate,
\begin{equation}
\av{\widetilde{m_i}}=\av{\frac{1}{N}\sum_{j=1}^N \xi_j^i}=\frac{1}{N}\sum_{j=1}^N \av{\xi_j^i}=\frac{1}{N}\sum_{j=1}^N \av{\xi^i}=m_i.
\end{equation}
This means that Eq.~(\ref{momest}) gives an \textit{unbiased estimate} of the true value of the moment, $m_i$.

The situation is different for the excess kurtosis. Consider an estimate similar to Eq.~(\ref{momest}),
\begin{eqnarray}
	\overline{\kappa}&=&\frac{1}{N}\sum_{j=1}^N \kappa_j=\frac{1}{N}\sum_{j=1}^N 
	\frac{\widetilde{m_{4,j}}}{\widetilde{m_{2,j}}^2}-3 \nonumber \\
	&=&\frac{1}{N}\sum_{j=1}^N \frac{m_4+\delta m_{4,j}}{(m_2
		+\delta m_{2,j})^2}-3 \,,
\label{kurtbiased}
\end{eqnarray}
where the deviations $\delta m_{i,j}$ of the measured moments $\widetilde{m_{i,j}}$ from their true values $m_{i,j}$ are introduced. These measured moments are themselves calculated from a finite series of samples using Eq.~(\ref{momest}) (of course, generally speaking, with a different number of samples $N$), and since these are unbiased estimates, $\av{\delta m_{i,j}}=0$. Assuming for simplicity that these deviations are small and expanding in a Taylor series up to second order, we get
\begin{equation}
	\overline{\kappa}\approx \frac{m_4}{m_2^2} \times\frac{1}{N}\sum_{j=1}^N 
		\left(1+\frac{\delta m_{4,j}}{m_4}\right)\left(1-2\frac{\delta m_{2,j}}{m_2}
		+ 3 \frac{\delta m_{2,j}^2}{m_2^2}\right) - 3 \,,
\label{kurtdev}
\end{equation}
which after averaging gives
\begin{equation}
\av{\overline{\kappa}}\approx \frac{m_4}{m_2^2}\left(1+3\frac{\av{\delta m_2^2}}{m_2^2}-2\frac{\av{\delta m_2\delta m_4}}{m_2 m_4}\right)-3 \ne \kappa.
\label{kurtbias}
\end{equation}
Thus, averaging the values of the excess kurtosis obtained from runs of finite duration will not lead to the correct value of $\kappa$ even in the limit of an infinite number of such runs.

A much better procedure is to average the moments over all runs and then use these averaged moments to estimate $\kappa$. That is, defining
\begin{equation}
\overline{m_2}=\frac{1}{N}\sum_{j=1}^N m_{2,j}
\end{equation}
and
\begin{equation}
\overline{m_4}=\frac{1}{N}\sum_{j=1}^N m_{4,j},
\end{equation}
we estimate $\kappa$ as
\begin{equation}
\widetilde{\kappa}=\frac{\overline{m_4}}{(\overline{m_2})^2}.\label{kurtunb}
\end{equation}
This is still, strictly speaking, not unbiased; the bias can be estimated using Eq.~(\ref{kurtbias}), with the deviations of the moments replaced by those of their averages ($\delta m_i \to \delta \overline{m_i}$). However, as $\av{\delta \overline{m_2}^2}$ and $\av{\delta \overline{m_2} \,\delta\overline{m_4}}$ both decrease as $1/N$ with growing $N$ [see Eqs.~\eqref{deltam22av} and \eqref{deltam2m4av} below], the bias not only vanishes as $N\to\infty$, but is also negligible compared to the uncertainty of the kurtosis (which goes as $N^{-1/2}$) once $N$ is large enough. Therefore, the estimate (\ref{kurtunb}) can be considered unbiased and we refer to it as such here and in the main text; it is this estimate that is used to produce the solid symbols in Fig.~4(a). Moreover, even when binning the data to produce the gray curves for individual runs in that plot, the same procedure is used to obtain each data point, although in this case the difference compared to using Eq.~(\ref{kurtbiased}) is very minor.

The uncertainty of the estimate (\ref{kurtunb}) can be estimated by using Eq.~(\ref{kurtdev}) with $N=1$, $m_i \to \overline{m_i}$ and $\delta m_i \to \delta \overline{m_i}$. This gives, to first order,
\begin{equation}
	\delta\widetilde{\kappa}\approx (\widetilde{\kappa}+3) \left(
	\frac{\delta \overline{m_4}}{\overline{m_4}}-2\frac{\delta \overline{m_2}}{\overline{m_2}}
	\right) \,,
\end{equation}
and then
\begin{equation}
	\av{\delta \widetilde{\kappa}^2}\approx (\widetilde{\kappa}+3)^2 
	\left( \frac{\av{\delta \overline{m_4}^2}}{\overline{m_4}^2}+4\frac{\av{\delta \overline{m_2}^2}}
	{\overline{m_2}^2}-4\frac{\av{\delta \overline{m_2}\delta\overline{m_4}}}
	{\overline{m_2}\overline{m_4}} \right) \,.
\end{equation}
The square root of the last expression, where the uncertainties of the averages are estimated as
\begin{eqnarray}
	\av{\delta \overline{m_2}^2}&\approx & \frac{1}{N(N-1)} \sum_{j=1}^N (m_{2,j}-
	\overline{m_2})^2, \label{deltam22av} \\
	\av{\delta \overline{m_4}^2}&\approx & \frac{1}{N(N-1)} \sum_{j=1}^N (m_{4,j}-
	\overline{m_4})^2,\\ 
	\av{\delta \overline{m_2}\delta \overline{m_4}}&\approx & \nonumber \\
	& &\hspace{-1.7cm}\frac{1}{N(N-1)}\sum_{j=1}^N (m_{2,j}-\overline{m_2})(m_{4,j}
	-\overline{m_4}) \,,
\label{deltam2m4av}
\end{eqnarray}
is plotted as the error bars in Fig.~4(a) in the main text.

\section{Critical time interval}

We have seen that for short time intervals $\Delta t$, the displacement distribution $P(\Delta z ;\Delta t)$ is dominated by the diffusing diffusivity effect, whereas at longer time intervals, it is dominated by the shape of the potential $U(z)$.  In this section, we derive the scale value $\Delta t_c$ that divides the two regimes.  As stated in the main text, the basic idea is to balance the diffusion and drift terms in Eq.~(3) of the main text.  That is, we want 
\begin{align}
	v_d^{(0)} \, \Delta t_c \approx \sqrt{2D_\perp^{(0)} \Delta t_c} \,,  \quad \implies \quad
	\Delta t_c \approx \frac{2D_\perp^{(0)}}{\left[v_d^{(0)}\right]^2} \,,
\end{align}
where $v_d^{(0)}$ is, crudely speaking, the average or typical \textit{absolute value} of the drift velocity and, similarly, $D_\perp^{(0)}$ is the typical value of the vertical diffusivity.

Our first task is to find the most probable distance $z_0$ between bead and wall, which we do from the minimum of the potential, $U'(z_0) = 0$.  Using Eq.~(2) in the main text, we have, with $\bar{U}$ denoting energies in units of $k_BT$,
\begin{align}
	\bar{U}'(z) = -\frac{\bar{B}}{\ell_D} \, \mathrm{e}^{- z/\ell_D} + \frac{1}{\ell_g} = 0 \,,
\end{align}
which implies
\begin{align}
	z_0 = \ell_D \ln \left( \bar{B} \frac{\ell_g}{\ell_D} \right) \,.
\end{align}

The next task is to estimate the typical force exerted by the potential as the particle fluctuates about the minimum at $z_0$.  We do so by approximating the motion about the minimum $z_0$ as a harmonic trap with ``spring constant" $k = U''(z_0)$, given by
\begin{align}
	\bar{U}''(z_0) = + \frac{\bar{B}}{\ell_D^2} \, \mathrm{e}^{- {z_0}/{\ell_D}} 
		= \frac{1}{\ell_g \, \ell_D} \,,			
\end{align}

The equipartition theorem then implies a typical displacement
\begin{align}
	\delta z = \pm \sqrt{\frac{k_BT}{k}} = \pm \frac{1}{\sqrt{\bar{U}''(z_0)} } = \pm \sqrt{\ell_g \, \ell_D} \,.
\end{align}
and a typical force / $k_BT$ of
\begin{align}
	 \bar{U}''(z_0) \, \delta z = \pm \sqrt{\bar{U}''(z_0)} = \pm \frac{1}{\sqrt{\ell_g \, \ell_D}} \,.
\end{align}

The drift velocity has two contributions:   
\begin{align}
	v_d &\approx \underbrace{{D_\perp}'(z_0)}_{\rm variable~\it D} 
		- \underbrace{{D_\perp}(z_0) \, \left[ \bar{U}''(z_0) \, 
			\delta z \right]}_{\rm force~from~potential}  \nonumber \\[3pt]
	&\approx \frac{D_0}{a} \, \left( 1 \mp \frac{z_0}{\sqrt{\ell_g \, \ell_D}} \, \right) \,,
\end{align}	
where we have approximated the diffusivity and its derivatives by their values at $z_0$ and, since the beads stay close to the substrate, $D_\perp(z) \approx D_0 (z/a)$ and ${D_\perp}'(z) \approx D_0/a$. For the typical value $v_d^{(0)}$ we then take the average of the two absolute values (with the ``$+$'' and ``$-$'' signs), and since under our conditions $z_0/(\ell_g\ell_D)^{1/2}>1$, this gives
\begin{align}
v_d^{(0)}=\frac{D_0 z_0}{a\sqrt{\ell_g \, \ell_D}}\,.
\end{align}
Similarly for the diffusivity,
\begin{align}
D_\perp^{(0)}&=\left[D_{\perp}(z_0-[\ell_g\ell_D]^{1/2})+D_{\perp}(z_0+[\ell_g\ell_D]^{1/2})\right]/2\nonumber\\
&\approx D_{\perp}(z_0)\approx D_0\frac{z_0}{a}.
\end{align}

\begin{figure}[h]
	 \begin{center}
	 \includegraphics[width=3in]{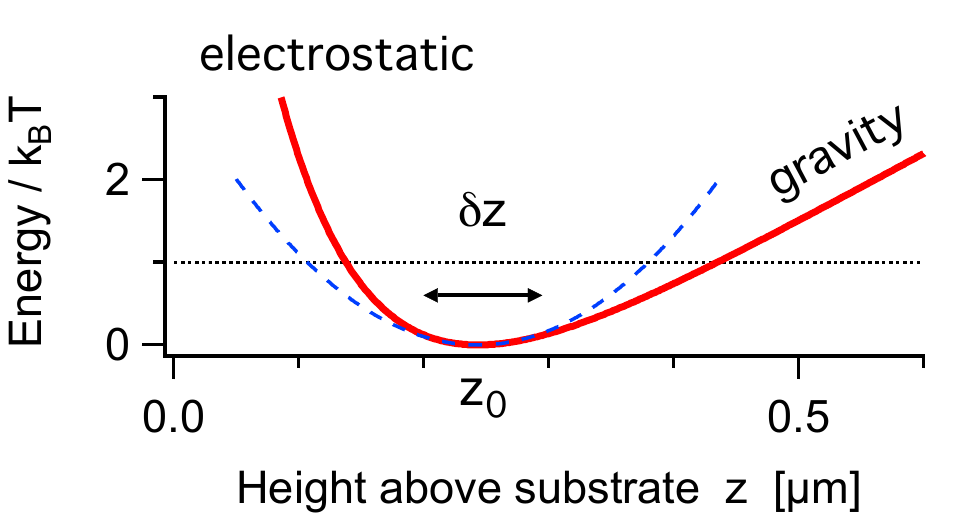}
	 \end{center}
	 \caption{(color online). Particle potential (thick solid curve), generated using parameters from Run 11.  The potential minimum, at $z_0$, divides the electrostatic and gravitational regimes.  The harmonic approximation is indicated by the dashed curve.  Dotted line is $k_BT$ above the minimum. The typical fluctuation scale is of order $\delta z$.}
\label{fig:potGraph}
\end{figure}

Finally, we compute the balance between drift and diffusion in this approximation.  In dimensionless units,
\begin{align}
	\overline{\Delta t_c} &= \Delta t_c \left( \frac{D_0}{\ell_g^2} \right)  \nonumber \\
	&= \left( \frac{D_0}{\ell_g^2} \right) \, \frac{2 D_\perp^{(0)}}{\left[v_d^{(0)}\right]^2} = 
		\frac{2a\ell_D}{z_0\ell_g}\nonumber \\[3pt]
	&= \frac{2a}{\ell_g \ln \left( \bar{B} \frac{\ell_g}{\ell_D} \right)}
		 \approx 12 \,.
\label{xover}
\end{align}
The last step uses numbers from Run 11:  $\ell_g = 0.123$ $\mu$m, $\ell_D = 0.072$ $\mu$m, $\bar{B} = 16.8$, and $a = 2.46$ $\mu$m.  These numbers imply $z_0 \approx$ 0.24 $\mu$m and $\delta z \approx$ 0.094 $\mu$m. The result is quite close to the experimental crossover at $ \overline{\Delta t_c} \approx 20$, given all the approximations involved and noting, in particular, that, as Fig.~\ref{fig:potGraph} shows, the harmonic approximation is not very accurate already at an energy $k_BT$ over the minimum.  Interestingly, the crossover between diffusion and Boltzmann regimes in the MSD plot of Fig.~4b gives a crossover at $\overline{\Delta t_c} \approx 13.6$, which is even closer to our estimate here.

It is also interesting to note that the same result (\ref{xover}) can be obtained in a different way, by considering the time it would take the particle to diffuse over the length of the characteristic interval, from $z_0-(\ell_g\ell_D)^{1/2}$ to $z_0+(\ell_g\ell_D)^{1/2}$. Since for most of the time interval $\Delta t_c$ diffusion dominates over drift, we neglect the drift altogether and then the typical time $\Delta t_c$ it takes to diffuse the distance $2(\ell_g\ell_D)^{1/2}$ is the solution of
\begin{align}
2\sqrt{\ell_g\ell_D}\approx\sqrt{2D_\perp(z_0)\Delta t_c}\,,
\end{align}
which gives (\ref{xover}).

\section{Simulations}

To confirm our understanding of the experimental results, we have carried out computer simulations of the motion of colloidal particles. The purpose is to verify that taking into account the factors described in the main text (namely, the electrostatic and gravity forces and the variation in the diffusivity with the height above the substrate) is sufficient to match the experiments qualitatively and quantitatively, while neglecting the diffusivity variation leads to significant discrepancies with the experimental results.

\begin{figure}[h]
	 \begin{center}
	 \includegraphics[width=3in]{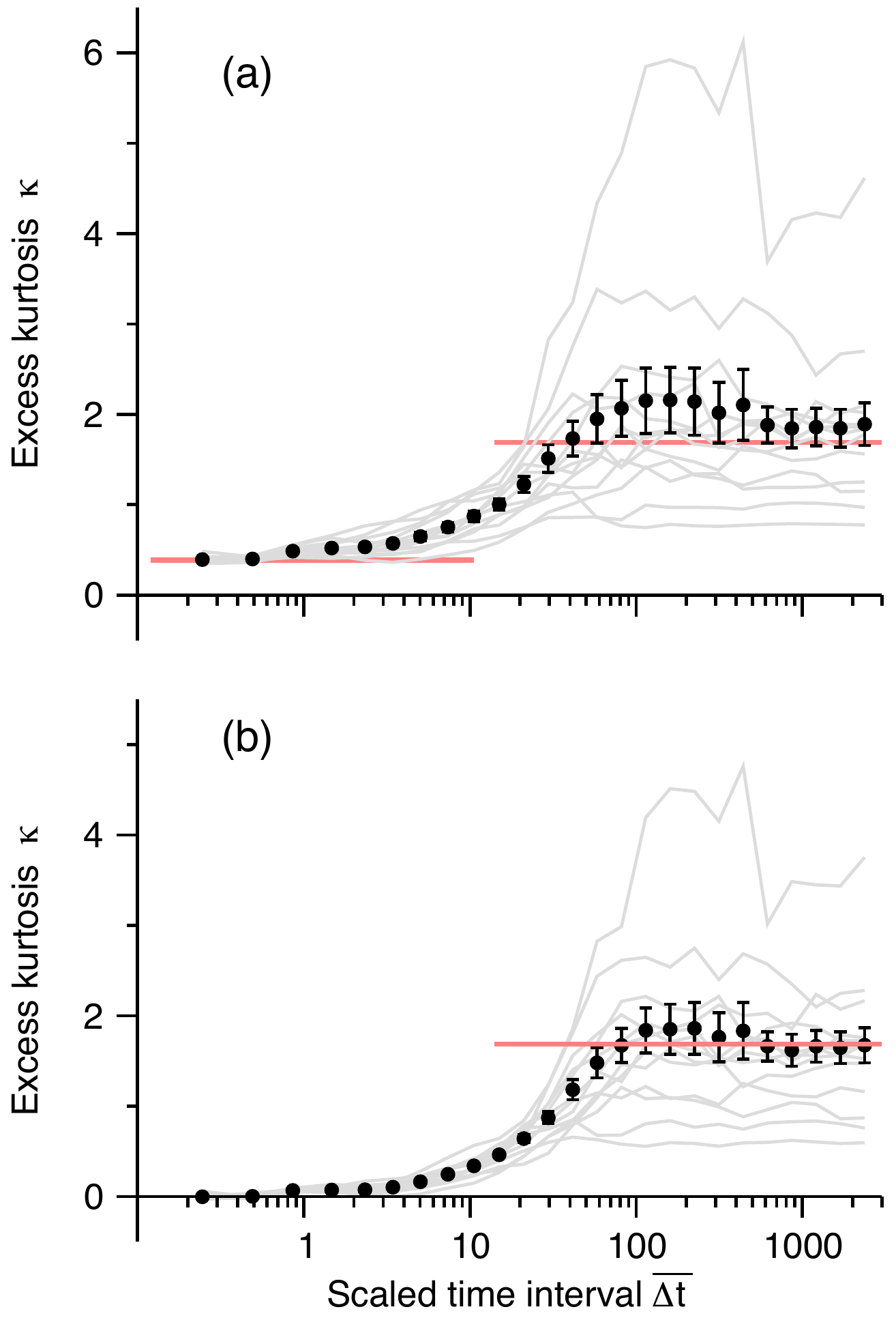}
	 \end{center}
	 \caption{(color online). Excess kurtosis of the vertical displacement distribution as a function of the time interval obtained in simulations, with the vertical diffusivity either (a) given by Eq.~(1) in the main text or (b)  constant and equal to the theoretical average.  All other parameters roughly match the experimental conditions.  The gray curves are the binned results for 13 individual realizations, and the solid circles with error bars are unbiased estimates based on these realizations.  The horizontal red lines are the theoretical short- and long-time-interval limits. For details, see the text.}
\label{fig:kurtsim}
\end{figure}

To simulate the motion of the particles, we have used the Brownian Dynamics approach. As in experiments, 13 runs were carried out. Since we are interested in the evolution of the height $z$ and the vertical motion is decoupled from the horizontal, a 1D simulation is sufficient. At time $t=0$ a particle starts at the height $z$ drawn from the Boltzmann distribution corresponding to the potential of Eq.~(2). Then at each simulation step $z$ changes according to Eq.~(3), with the time step $\delta t=0.001$~s. The diffusivity varies with height according to Eq.~(1). Since the experimental conditions vary little between the experimental runs, we have used the same parameters for all our runs, close to the average values in the experiment. Namely, $\ell_g=0.116\ \mu$m, $\ell_D=0.076\ \mu$m, $\overline{B}=16$, $a=2.51\ \mu m$, and $D_0=0.1\ \mu\rm{m^2/s}$. Data for the $z$ position were collected every 33 steps, or 0.033~s, the same interval as in the experiment. We have also done a series of runs emulating the ``motion blur'' effect [see Eq.~(\ref{eq:Dperp_est})], by averaging the position over 11 time steps. This has a very minor effect, which justifies neglecting this effect, e.g., when calculating the MSD and the excess kurtosis. These results are not presented here. Since one of our purposes is to verify that the significant discrepancies between different runs at large $\overline{\Delta t}$ seen in Fig.~4(a) are attributable to lack of statistics, we have used the same simulation durations as in the experimental run. For each run, the kurtosis of the displacement distribution was first calculated for time interval durations from 1 to 11249 0.033-second time steps; for each duration, the set of data used in the calculations consisted of the displacements over all the intervals of that duration (including overlapping ones) present in the run. To smooth the resulting kurtosis dependence, the data were binned, as in the experiments. The result is shown in Fig.~\ref{fig:kurtsim}(a), together with the unbiased estimate based on all 13 runs (solid circles with error bars).  There is good agreement, both qualitative and quantitative, with the experimental results in Fig.~4.

We have also repeated the same simulations with one significant difference: the diffusivity was made constant and equal to $D=0.0104 \ \mu {\rm m^2/s}$, the value obtained by averaging Eq.~(1) with the Boltzmann distribution. To facilitate the comparison with the variable-$D$ case, the same seed value was used for the pseudorandom number generator in both cases. While the results [Fig.~\ref{fig:kurtsim}(b)] are similar for large $\overline{\Delta t}$, they are very different for small $\Delta t$, with the kurtosis in the constant-$D$ case close to zero and significantly below the experimental values. The MSD dependences are nearly identical in the two cases and similar to Fig.~4(b) (not shown).

Simulations were also used to estimate the uncertainties of the fitting parameters in Table~\ref{ptable2}. For this, we carried out two series of runs (400 runs in each) with identical parameters (same values close to the averages as quoted above). In one series, the lengths of all runs were as in the shortest experimental run (Run 8), and in the other series, as in the longest run (Run 10). For each of the 800 runs, the position distributions were obtained using the same bin width as for the experimental data and fitted with the Boltzmann distribution. For an infinitely long run, the fitting parameters would be identical to the simulation parameters, but for runs of a finite duration the distributions have random noise and so do the parameters of the fits. The standard deviations of these parameters are then used as the estimates of the uncertainties of the parameters in the experimental fits.


\begin{thebibliography}{39}
\bibitem{uncert} The uncertainties of these parameters given here and in the main text are statistical and omit comparable systematic uncertainties due, for example, to temperature calibration. The statistical uncertainties are what is relevant when comparing measurements using different beads.
\end{thebibliography}
\end{document}